\documentclass[preprint,preprintnumbers,amsmath,amssymb,superscriptaddress,nofootinbib,,showpacs]{revtex4}

\usepackage{graphicx,amsfonts}
\usepackage{epsfig}
\usepackage{epstopdf}
\usepackage{dcolumn}
\usepackage{bm}
\usepackage{comment}
\usepackage[ansinew]{inputenc}
\usepackage{textcomp}
\begin{document}

\title{Semitauonic $B$ Decay Anomaly}

\author{ E. C. F. S. Fortes}
\affiliation{ Instituto de F\'\i sica Te\'orica, Universidade Estadual Paulista, Rua Dr. Bento Teobaldo Ferraz 271,
01140-070 S\~ao Paulo, SP, Brazil}

\author{Shmuel Nussinov}
\affiliation{ Raymond and Beverly Sackler School of Physics and Astronomy, Tel-Aviv University, Tel-Aviv 69978, Israel.}

\begin{abstract}
 The anomalously large experimentally measured ratios of the  semitauonic decay $B\rightarrow D^{(*)} +\tau+\nu$  and the corresponding semileptonic  $B\rightarrow D^{*} +\l+\bar{\nu}_l$  disagree with the predictions of the standard  E.W + QCD model(S.M).

We briefly comment on this disagreement and on possible new physics explanations which are rather constrained and difficult to implement.

\end{abstract}


\maketitle
\section{Introduction}

${}$
Searches for physics beyond the standard model (BSM) of particle physics are ongoing at high energy and high intensity, high precision experiments, and much theoretical effort is devoted to computing  the rates expected in the standard model (SM) or to BSM scenarios accommodating whatever anomalies exist at the time and also dark matter and neutrino masses/mixing.
 Many ``anomalies" arose over the last decades, but no single clear conflict with the SM presently exists (and this includes the discovery of the Higgs particle at the LHC). Some anomalies such as the high energy $t-\bar{t}$ asymmetry at  Fermi-lab were not supported by the higher energy LHC and by further calculations of the SM expected values.
In this context the excess in semitaunic $B$ meson decays seems to be rather unique.
It has been seen in $B\rightarrow D^{(*)}+\tau+\bar{\nu}_{\tau}$ by both high intensity BABAR\cite{Lees:2012xj, Lees:2013uzd} and Belle\cite{Huschle:2015rga} electron positron colliders, and the $B\rightarrow D^*$ excess was recently confirmed by LHCb\cite{Aaij:2015yra} in a $10^3$ times higher energy proton collider.
Also the heavy quark effective field theory (HQEFT) developed over the last decades\cite{Manohar:2000dt} reliably predicts the differential rates for these  and other  $B$ decays in the (QCD corrected) SM . In particular the $B\rightarrow D^{(*)}$ form factors can be extracted from the measured  $B\rightarrow D^{(*)}+l+\bar{\nu_{l}}$ with $l =e$  or $\mu$ to predict the tauonic decay rates with the HQEFT fixing the form factors in terms of few model parameters\cite{Kowalewski:2008zz}.

The above and the prospect of more LHCb results for decays of non-$B_{u,d }$ meson $b$ flavored hadrons prompt us to further consider this anomaly.
 Recently we became aware of the paper \cite{Freytsis:2015qca} that addresses the issues  with greater precision depth and detail and also provides an extensive  bibliography and we will keep referring to this paper as we go along.  In view of the importance of the general subject , the simpler version of the inclusive decay rates that we present and the different BSM model involving a charged uncolored new exchange (rather than a leptoquark exchange)  that we emphasize,  our present paper may still be useful.

In the next section we present a crude yet transparent estimate suggesting that the SM is unlikely to reproduce the large measured ratios $R_D$ and $R_{D^*}$ of the semitaunic and semileptonic modes. In Sec. III we comment on explanations of the anomaly using new particles exchanged in the various channels of the semitaunic process.  In Sec. IV we focus on a charged uncolored $X^{-}$ particle mediating only $2\rightarrow 3$ generation mixing.

\section{ Simple Inclusive Approach to Estimating the Semitauonic Decays}
${}$
We do not attempt to reproduce the accurate SM calculations of the ratios of semitauonic and semileptonic $B$ decays $R_D^{(*)}=\Gamma (B$
$ \rightarrow D^{(*)}+\tau+\bar{\nu}_\tau)/\Gamma (B\rightarrow D^{(*)}+l+\bar{\nu_{l}})$. We believe that these calculations, building on the HQEFT, are correct, though further verification by using presently much advanced lattice calculations would clearly be of great importance.
It is easy to see why at the nonrecoil point the $B\rightarrow D(D^{*})$ semileptonic transition becomes universal if the $b$ and $c$ quarks were infinitely heavy. For $\overrightarrow{p}(c)=-[ \overrightarrow{p}(l)+\overrightarrow{p}(\nu)] =0$, the $b \rightarrow c $ transition changes nothing for the ``spectator quark" or other light degrees of freedom of QCD which still see the same static color charge at the origin, possibly with a flipped spin. This does not change energies in this limit and the relevant form factor becomes 1.
However, to extend beyond this special point using a systematic expansion in $\Lambda(QCD)/{m(c)}$, one needs the complete machinery of HQEFT. While briefly described in particle data group (PDG) minireviews, most high energy physicists may find the actual calculations of $R(D^{(*)})$ shrouded in technical details and, like other monumental calculations of lattice QCD, simply take them on faith. Because of their extreme importance, the following crude yet easy to follow estimate of the $R$ ratios may be of some use.
Rather than addressing separately the $D$ and $D^*$ cases, we will consider the ratio of the sum of the two modes in the semileptonic and semitauonic cases:
\begin{equation}\label{a}
  R( D + D^{*})= \frac{\Gamma(B\rightarrow D+ \tau+ \bar{\nu}_{\tau}) + \Gamma(B\rightarrow D^*+ \tau+ \bar{\nu}_{\tau})}{[\Gamma(B\rightarrow D +l +\bar{\nu}_{1})]+ [\Gamma(B\rightarrow D^{*} +l + \bar{\nu}_{l})]}.
\end{equation}

   Consistent values of the CKM matrix element$ V_{b,c}$ were obtained by analyzing the exclusive $D$ and $D^*$ decay channels and independently via the inclusive semileptonic decay.
The latter calculation used quark hadron duality formally referred to as the short distance expansion. It is based on the separation of time scales between the fast quark level $b\rightarrow c +l+\bar{\nu}_l$ transition of duration $\delta(t) = 1/(m_b-m_c) $ and the longer stage lasting $\Delta(t)= 1/{\Lambda(QCD)}$ during which the charmed quark combines with the spectator  $\overline{d}$ to form the final hadronic state . The hadronization does not affect the total inclusive decay rate which is modified, however from the zeroth order, partonic calculation by QCD corrections. These were computed to various orders and included in the above-mentioned work of Freytsis-Ligeti and Ruderman \cite{Freytsis:2015qca}. While these authors also used the inclusive calculation as further and less model dependent evidence, for the tension between SM calculation and the experimental measured $R(D)$ and $R(D^{*})$, our approach is simpler and different.

The idea is that the inclusive $B\rightarrow X_{c}+ \tau+ \overline{\nu}_{\tau}$  (where $X_c$ is a charmed hadron) or rather the ratio
\begin{equation}\label{b}
  R(\rm Inclusive) = \Gamma(B\rightarrow X_c+ \tau+ \bar{\nu}_\tau)/\Gamma(B\rightarrow X_c +l + \bar{\nu_{l}}),
\end{equation}
can provide an upper bound on  $R(D+D^*)$. The bound is obtained by assuming that, in the $\tau$ case, the complete inclusive decay is channeled into the $D$ and $D^*$ final states only.

\begin{figure}[t]
  \centering
  \includegraphics[width=11cm]{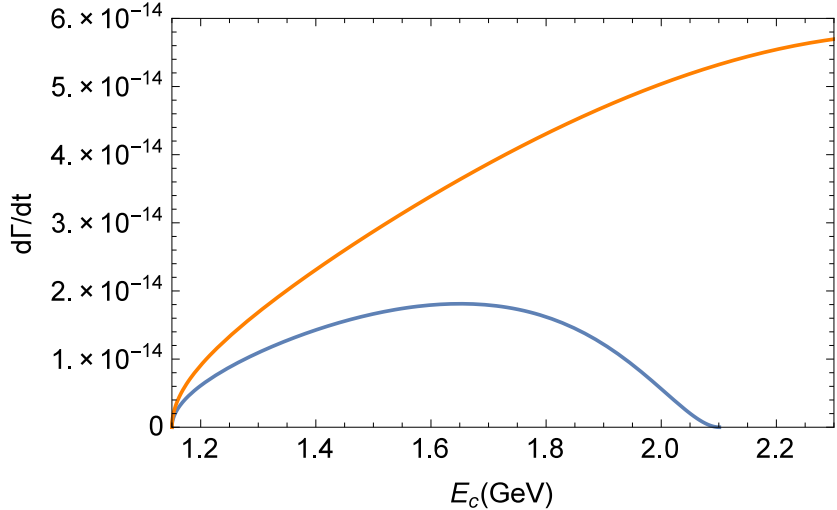}\\
  \caption{Differential decay width of the $b$ quark  as a function of $E_{c}$. The color orange represents the leptonic  $l=e,\mu$ cases and the blue one represents the tauonic decays.}\label{1b}
\end{figure}

The main factor fixing $R^{0}$(Inclusive), where the zero index denotes the zeroth order partonic calculation of $R(\rm Inclusive)$, is phase space which is significantly smaller for the semitauonic decay.
$E_c$ or $t = (p_{b}-p_{c})^2=[ m_{b}^2+m_{c}^2- 2m_{b}E_c]$ is the only kinematic variable relevant for the hadronization. Lower $E_c$ i.e. smaller $c$ recoil momenta  favors the formation of the ground states $D$ and $D^*$. As evident from Fig. \ref{1b}, the  distribution of $E(c)$ in the decays to final states with the heavier $\tau$, peaks at lower values than in the $l=e,\mu$ case. This suggests a larger fraction of $D$ and $D^*$ in the inclusive  tauonic decays (we have used the values of $m_b= 4.4$ GeV and $m_c=1.275$ GeV taken from Ref.\cite{Benson:2003kp}).

The ratio between  inclusive semitauonic and semileptonic $B^0$ decays for the above $m_b$ and $m_c$ was very small, $R^{0} (\rm Inclusive) \sim 0.14 $.
 Motivated by the comments at the end of the previous paragraph, we make the extreme assumption that the semitauonic inclusive decay goes \textit{only} into the $D$ and $D^*$ final states.
This \textit{is not} the case for the other semileptonic $B$ decays where from the PDG we find that only $\sim$ 70\% of the decays are into $D$ or $D^*$. With the remaining 30\% corresponding to higher $D^{**}$, nonresonant $D^{(*)} + n \pi $ or even baryonic $\Lambda_c + N $ states. Since 100\% of all the inclusive semitauonic decays go into $D$ or $D^*$, this  enhances $R(D+D^*)$  to $0.2 $ which is still much less than $R(D+D^{*})_{\rm experimental}\simeq  0.34$.
 To test for the sensitivity of the results to $m_b$ and $m_c$, we varied $m_b$ between $4.4$ and $4.6$ GeV and $m_c$ between $1.15$ and $1.275$  GeV. We obtained $R^{0}(\rm Inclusive)$ values ranging from $0.14$ to $0.234$ and corresponding maximal $R(D+D^{*})$ between $0.2$ and $0.335$. Except for the  last case, all the $R(D+D^{*})$ are in disagreement with the experimental values.

 We note a certain part of the $O(\alpha_{s})$ corrections is particularly pertinent to the present approach,  further decreases $R(\rm Inclusive)$.  In addition to the $b\rightarrow c+l+\bar{\nu}_l$ process, we have the distinct four body perturbative final state,  where an extra gluon has been radiated, $b\rightarrow c+l+\bar{\nu}_l + g$. We estimate that this enhancement is $\sim  10\%$ of the total inclusive decay.
Due to  the much more dramatic decrease of the four body $c+\tau+\bar{\nu}_{\tau} + g$  phase space with increasing lepton masses, we expect that the latter part is almost completely absent in the inclusive $B^{0}$ decay to charm  and the massive $\tau$  lepton. Thereby decreasing $R(\rm Inclusive)$ and also the maximal $R(D+D^{*})$ by up to an extra 10\%.

\section{Possible Explanations of the Large $R(D)$ and $R(D^{*})$}
${}$
A new anomaly which appear only when all the quarks and leptons involved are members of the third generation is attractive in the framework of extradimensional theories allowing partial compositeness but only for third generation fermions\cite{Agashe:2006wa}. The present anomaly involves beside the $b$, $\tau$, and $\nu_{\tau}$ also the $c$ quark which up to the $V_{bc}\simeq 0.04$ mixing is a member of the second generation. Still it comes closer to the ideal third generation fermions only than all previously studied cases, in particular the $B_{u}\rightarrow\tau \nu_{\tau}$ decay involving a first generation $u$ quark and in which no anomaly was reported.

Existing and forthcoming lower bounds on masses of new particles and/or on deviations from the SM in other processes, constrain new physics interpretations of the $R(D)$, $R(D^{*})$ anomaly. We next proceed with a general phenomenological discussion (which is less detailed and somewhat different from that of Ref. \cite{Freytsis:2015qca} ). The anomaly can be due to a new particle $X$ which interacts with the $b$ and $c$ quarks , $\tau$ and $\bar{\nu}_{\tau}$ leptons.
\textit{A priori} this can happen in three distinct ways:
\begin{itemize}
  \item The $X$ particle could be color and charge neutral and exchanged between the $b$ and $c$ quarks and between the $\nu_\tau$ and $\tau$ (case a).
  \item The $X$ particle could be a leptoquark exchanging between the $b-\nu_{\tau}$ and a $c-\tau$ vertex or between a $b-\tau$ and a $c-\overline{\nu}_\tau$ vertex (case b).
  \item The $X^{-}$ particle could be exchanged just like the ordinary SM $W^{-}$ in the $(b-c)$ or $\tau -\bar{\nu}_{\tau}$, i.e. the $t$ channel (case c).
\end{itemize}
In  case a the $X$ particle exchange is a new radiative correction of the ordinary weak current hadronic vertex and of the weak leptonic vertex.
 To explain the big $R(D)$  and $R(D^{*})$, these radiative corrections should be large, exceeding the strong QCD corrections of the hadronic vertex due to a gluon exchange. This new exchange should not occur between  the first family generation. This scenario is strongly constrained by the well-established lepton universality in $Z$ decays. A modified $\tau-\nu_{\tau}$ vertex will also change the rate of tau decays. Thus,  attributing the observed anomaly a diagonal $X$ coupling, which otherwise minimally affects the flavor structure of the theory, is practically excluded.

Case b, with the leptoquark $X=Lq$  particle, was discussed in great detail in Ref. \cite{Freytsis:2015qca} which found it to be the most likely explanation. As correctly pointed by the authors, this would require relatively light -O(TeV) $Lq's$ and the discovery of such particles in future LHC experiments would be sensational \footnote{Clearly discovering this quark-lepton unifying theme is of extreme importance independently of any detailed motivation. The leptoquark proposed by Ref.\cite{Freytsis:2015qca} is specifically a third generation type leptoquark which evades the very strong lower bounds on the masses of the putative first generation of leptoquarks and is presently limited by the LHC to be more massive than 0.5 TeV only. Heavy  $10^{15}$ GeV leptoquarks  first arose in the framework of grand unified theories. It will be very remarkable if the eventual discovery of a far lighter leptoquark will be motivated by the much more humble $b$ flavor physics.}.

Here we consider case c, namely the possibility of a $t$ channel exchange of a new charged, uncolored $X$ particle.
This was addressed before in particular by e.g. Refs \cite{Tanaka:1994ay} and \cite{Fajfer:2012sv}. It is known  \cite{Kowalewski:2008zz} that a charged scalar particle in the  two Higgs doublet (2HDM) type II extensions of the SM  fails to reproduce \textit{both} $R(D)$ and $R(D^*)$ for any combination of masses and couplings. On the other hand, in Ref.\cite{Crivellin:2012ye,Crivellin:2015hha} it was shown that it is possible to fit $R(D)$ and $R(D^*)$ with to the 2HDM type III with some set of parameters. Reference \cite{Celis:2012dk} suggested that  the aligned 2HDM  can fit both $R(D)$ and $R(D^*)$, but the ranges of parameters used to do this fit are in conflict with the constraints from leptonic charm decays, so they analyze the predictions of the model for observables ($\tau$ spin asymmetry, forward-backward asymmetries, and differential decay rates)  sensitive to the charged-scalar contributions.

Here we  focus on a vector $X$ particle with $V-A$ couplings to quarks and leptons. As noted by Ref.\cite{Freytsis:2015qca} this simplifies the analysis since the new BSM amplitude has  the same form as the ordinary $W$ exchange and generates the same four-Fermi effective Lagrangian with a new Fermi constant,
\begin{equation}\label{n1}
  G_{X}=\frac{g_{X}^{2}}{M_{X}^{2}}\frac{M_{W}^{2}}{g_{W}^{2}}G_{Fermi}
\end{equation}
The Large Electron Positron Collier (LEP) bound ( $M_{X^{-}}> 80$ GeV from pair productions) is likely to be improved by the LHC. It is important to note that  $X^{-}X^{+}$ pair  production rates are smaller than those of pairs of leptoquarks ($Lq$) by a factor of $(\alpha_{em} /\alpha_{s})^{2} \simeq 7\times 10^{-3}$. Therefore, we expect that the new bounds on $M_{X^{-}}$ which at present are of order 200 GeV will be significantly weaker than those for leptoquarks. This may allow such particles to play the large role required to explain $R(D+D^{*})$ and  is an important motivation to consider the following toy model.

 \section{Toy Model Based on the Exchange of a Vector $X^{-}$ Particle For The  $R(D)$, $R(D^{*})$ Anomaly }
 ${}$
 We next proceed to discus  a toy model involving second and third generations only, which is based on the exchange of a charged, color neutral, vector $X^-$ particle in the $t$ ( $b\bar{c} \rightarrow \tau\bar{\nu}_{\tau}$) channel.
 Many BSM extensions modifying the flavor structure lead to excessive FCNC transitions and in particular to large $B_{s}-\overline{B_{s}}$ and $D-\overline{D}$ mixing.  To avoid an $X$ particle tree-level exchange from generating this most acute mixing problem, we endow our $X's$ particles with conserved generation numbers. Specifically we denote the required $X$ boson as $_{\textrm{2u}}X_{\textrm{3d}}$. This refers to the second and third generations each including a quark and lepton $u-d$ doublets. Thus, we denote $t=3u|q$, $b=3d|q$, $\nu_{\tau}=3u|l$ and $\tau=3d|l$ and also $c=2u|q$, $s=2d|q$, $\nu_{\mu}=2u|l$, and $\mu=2d|l$. Our proposed $X$ particle can generate the quark $b\rightarrow c$ transition and its conjugate can generate the leptonic $\nu_{\mu}\rightarrow \tau$ transitions. Thus, in our model the new BSM contribution involves a different final state $\overline{\nu}_{\mu}$ rather than $\overline{\nu}_{\tau}$ which is emitted in standard model $W$ exchange. Clearly the experiments where the $R(D)-R(D^{*})$ anomalies were discovered cannot distinguish neutrinos with different flavors. However, an important consequence is that in the present approach the SM and BSM amplitudes do not interfere and the ratio of their contributions is
 \begin{equation}\label{a1}
   \frac{G_{X}^{2}}{G_{F}^{2}}=\left[\frac{(g_{X}/M_{X})}{(g_{W}/M_{W})}\right]^{4},
 \end{equation}

so that $\Gamma_{X}(b\rightarrow c+\tau+\overline{\nu}_{\mu})/\Gamma_{(W=SM)}(b\rightarrow c+\tau+\overline{\nu}_{\tau})=[(g_{X}/M_{X})/(g_{W}\sqrt{V_{bc}}/M_{W})]^{4}$. In order to explain the observed anomaly, the last ratio should be about $\approx 1/4$, and finally we obtain the condition
\begin{equation}\label{b1}
  \frac{g_{X}}{M_{X^{-}}}\approx 0.14 \left(\frac{g_{W}}{M_{W}}\right).
\end{equation}
  The fact that the present  lower bounds on $M_{X^{-}}$ are very low is very helpful in satisfying this constraint, e.g  by choosing  $M_{X^{-}}= 200$ GeV and $g_{X} = (1/3) g_{W}$.

The weak $SU(2)_{L}^{W}$ symmetry that should be respected implies that our charged $_{\textrm{2u}}X_{\textrm{3d}}$ is accompanied by a neutral $(_{\textrm{2d}}X_{\textrm{3d}}- _{\textrm{2u}}X_{\textrm{3d}})/\sqrt{2}$.  This neutral $X^{0}$ particle is the middle member of an $SU(3)_{L}$ triplet. Its third member is $_{\textrm{2u}}X_{\textrm{3d}}$,
\begin{eqnarray}
\left( \begin{array}{c}
   _{\textrm{3u}}X_{\textrm{2d}}^{+} \\
  \frac{ _{\textrm{2d}}X_{\textrm{3d}}^{0}}{\sqrt{2}}- \frac{ _{\textrm{2u}}X_{\textrm{3u}}^{0}}{\sqrt{2}}\\
   _{\textrm{3d}}X_{\textrm{2u}}^{-}
 \end{array}\right)
\end{eqnarray}
The linear combination $(_{\textrm{2d}}X_{\textrm{3d}}+ _{\textrm{2u}}X_{\textrm{3d}})/\sqrt{2}$ is orthogonal to the middle member of the triplet and is a singlet of weak isospin.

A key point is that the neutral $X^{0}$ do not mediate  $B_{s}-\overline{B}_{s}$ mixing  as can be readily verified by following the 2,3 generation indices. However, they do generate FCNC transitions.
In particular the neutral member of the triplet $ _{\textrm{2d}}X_{\textrm{3d}}^{0}/\sqrt{2}-  _{\textrm{2u}}X_{\textrm{3u}}^{0}/\sqrt{2}$ will generate both $b\rightarrow s+\overline{\nu}_{\tau}+\nu_{\mu}$ and the dramatic flavor changing $b\rightarrow s+\tau^{+}+\mu^{-}$ transitions which will be $\propto g_{X}^{4}/M_{X^{0}}^{4}$. Using the PDG  upper bound  $\sim 5\times 10^{-5}$ on the branching of $B\rightarrow K\nu\nu $ and of $B\rightarrow K^{*}\nu\nu$, we estimate the inclusive $b\rightarrow s+ \overline{\nu}_{\tau}+\nu_{\mu}$ to be less than $5\times 10^{-4}$. This implies the constraint
\begin{equation}\label{c1}
 \frac{ M_{W}}{M_{X^{0}}}\frac{g_{X}}{g_{W}}\leq 0.05.
\end{equation}
Notice that once  $M_{X^{0}}\geq 3M_{X^{-}}$, the last condition is satisfied. Further, if the bound on $b\rightarrow s+\overline{\nu}_{\tau}+\nu_{\mu}$ is saturated  we expect a similar slightly smaller rate due to the phase space reduction for the inclusive $b\rightarrow s+\tau^{+}+\mu^{-}$  which should be experimentally looked for.
Thus we predict an inclusive decay rate $\Gamma (B\rightarrow s+\tau^{+}+\mu^{-})=\Gamma(B\rightarrow K,K^{*},K^{**}, \texttt{etc})+\tau^{+}+\mu^{-}$ which is proportional to $g_{X}^{4}/M_{X^{-}}^{4}$. Weak isospin symmetry implies the same $g_{X}$ as in the previous equations.

Up to now we have considered the minimal set of extra particles that $SU(2)_L^{W}$ requires once we have the  $_{d3}X_{u2}$ boson mediating the BSM contribution explaining the observed excess of the $b\rightarrow c +\tau +\bar{\nu}$ . While we imposed the necessary equality of the couplings of these we had to assume a mass ratio of $\sim3$ the neutral to charged member of the resulting $SU(2)_L$ triplet.

One may want to extend the set of new particles by imposing a $U(2)$ family symmetry and use the $I_{2\rightarrow 3}$  generator operating on our $_{d3}X_{u2}$ boson to predict two new particles of the form $_{d3} X_{u3}$ $\pm _{d2}X_ {u2}$ . These generation diagonal bosons of mass $M_{X^{\prime -}}$ do not generate new FCNC transitions. However, they add a coherent contribution to the charged $\tau\rightarrow\nu_{\tau} +\mu^{-} + \bar{\nu}_{\mu}$ of size
$A_{sm}\cdot G_{X}/G_{F}\cdot M_{X^{-}}^{2}/M_{X'^{-}}^{2}$ with $A_{sm}$ the SM,  $W^{-}$ exchange contribution.  By assumption there is no analog contribution to the leptonic $\tau$ decay to electron and neutrinos.   Using the ratio $G_{X}/G_F\sim  0.02$ required to fit the $R$ anomaly, we find that a ratio of $\sim 2$ of the masses  is enough to avoid violation of the relatively well experimentally verified lepton universality of the $\tau$ decay in the first and second generation of families.

\begin{figure}[h]
  \centering
  \includegraphics[width=14.61cm]{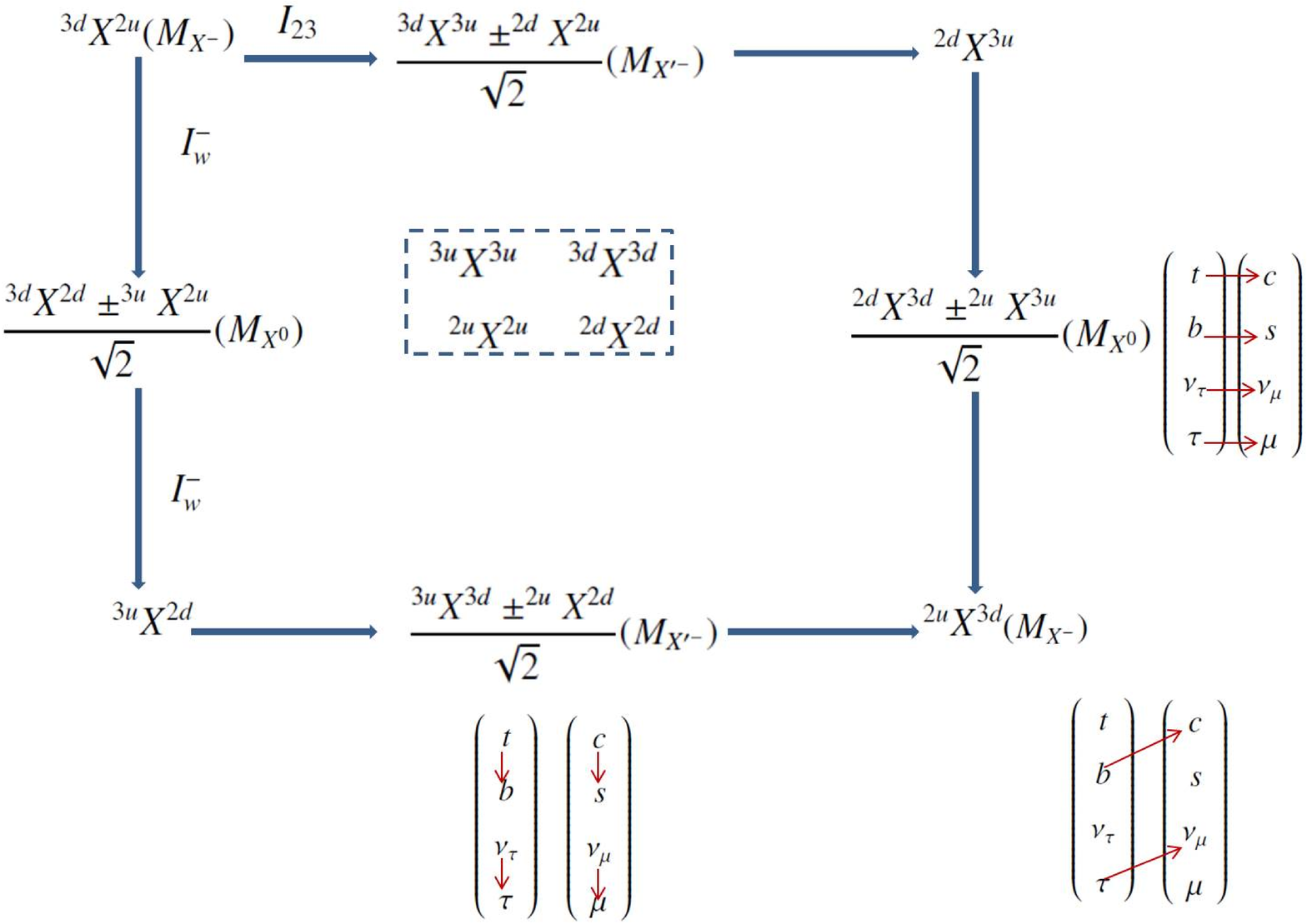}\\
  \caption{The 16 states corresponding to the various $_{23,ud}X_{23,ud}$ combinations.
 The particle at the upper left corner  of mass $M_{X^-}$  is the one we need to mediate the extra incoherent contribution $b\rightarrow c+\tau^- +\nu_{\mu}$.
 The $X^0$ particle of mass $M_{X^{0}}$ at the middle of the left edge is obtained from the initial $X^-$ by a weak isospin lowering operation. To each particle corresponds a conjugate particle mediating the reverse reaction  obtained by a   $ u \leftrightarrow d$ and $2\leftrightarrow 3$ reflection through the origin of the square. As indicated by the arrows corresponding to $X$ emission and absorption,  the $X^0$ particle mediates the decays $b\rightarrow s+\nu+\nu$ and $b\rightarrow s _\tau^{-} +\mu^{+}$. The $X'^{-}$ particle of mass $M_{X'^-}$ in the top middle was obtained by operation with the generation shifting operator $I_{2,3}$. It adds coherently to the standard model weak decay $\tau\rightarrow\nu_{\tau} +\mu^{-} + \bar{\nu}_{\mu}$ but not to the corresponding electronic tau decay which puts a lower limit on $M_{X'^{-}}$.}\label{fd1}
\end{figure}

 For completeness we indicate in Fig \ref{fd1} the  set of 16 bosons of the form $_{u,d 2,3} X _{u,d 2,3}$   that are generated by repeating  operations like the above and which form a $U(2)^{W} \times U(2)^{Gen}$ $4 \times 4$ representation.   As indicated no new types of transitions are then generated except for the completely (generation and weak isospin) diagonal transition due to the 4 mesons, which we refer to as $X^{0''}$, in the center of our figure. Since the latter are of the same type as those considered in case a, we need to impose on their masses an even stronger constraint than above namely
 $M(X^{0''}) \ge{2}$  TeV.

So far we have only addressed possible difficulties associated with the newly introduced $X$ particle sector on its own. To avoid difficulties with other precision measurements we had to impose a 1:3:6:10 hierarch between the masses of the required $X^-$ , its weak isospin $X^0$ partner, the generation neutral $X^{' -}$ and the completely generation and $u-d$ diagonal $X^{''0}$.  It seems difficult  to explain the large $SU(2)_L$ breaking manifest via the $X^{-} - X^{0}$ mass difference\footnote{ The large $t-b$ mass difference does not violate $SU(2)_L$ because it originates from Higgs coupling between left and right quarks where the latter are $SU(2)_L$ singlets. To have a remotely related scenario here we should involve also right-handed $X$ particles as well.}.

 We also need to address the all important issue of the breaking of flavor symmetries by the Higgs couplings. These were  reflected in the quark and lepton spurions in the paper of Ref.\cite{Freytsis:2015qca} . While it is conceivable that our model when fully implemented with its own new Higgs sector will be able to generate the magic of GIM-like mechanism avoiding  \textit{all}  the possible dangerous resulting FCNC, this seems rather unlikely. Discussing this alternative, less dramatic  than the leptoquark  explanation, is still very useful \cite{Freytsis:2015qca}.

  \section{Summary and Conclusions}

The $R(D)$ and $R(D^{*})$ which \cite{Freytsis:2015qca} and  we discussed  are very tantalizing.
  If further LHCb and Belle-2 experiments will verify the high ratios and exclude a very subtle systematic error plaguing all existing experiments, we will have to incorporate the anomaly into the theoretical BSM framework. The fact that it seems almost impossible to do so makes it only all the more challenging.

\section*{Acknowledgments}

  The authors thanks Andreas Crivellin and Martin Jung for useful comments. E.C.F.S.F. thanks University of Maryland and NASA Goddard Space Flight Center for the hospitality while this work was being completed and FAPESP for full support under Contracts No. 14/05505-6 and 11/21945-8.

\end{document}